\documentclass[lettersize,journal]{IEEEtran}
\usepackage{amsmath,amsfonts}
\usepackage{algorithmic}
\usepackage{algorithm}
\usepackage{array}
\usepackage[caption=false,font=normalsize,labelfont=sf,textfont=sf]{subfig}
\usepackage{textcomp}
\usepackage{stfloats}
\usepackage{url}
\usepackage{verbatim}
\usepackage{graphicx}
\usepackage{cite}
\usepackage{booktabs}
\usepackage{hhline}
\usepackage{hyperref}
\usepackage{multirow}
\usepackage{bbding}
\usepackage{amsthm}
\usepackage{cite}
\usepackage{amsmath,amssymb,amsfonts}
\usepackage{inconsolata}
\usepackage{xcolor}
\usepackage{threeparttable}
\usepackage{framed}
\usepackage{multicol}
\usepackage{mdwlist}

\usepackage{enumitem}

\definecolor{BrickRed}{RGB}{178,34,34}

\makeatletter  
\newif\if@restonecol  
\makeatother

\usepackage[
  n,
  operators,
  advantage,
  sets,
  adversary,
  landau,
  probability,
  notions,  
  ff,
  mm,
  primitives,
  events,
  complexity,
  asymptotics,
  keys]{cryptocode}

\begin{document}
\title{AutoIoT: Automated IoT Platform Using Large Language Models}

	\author{
 Ye Cheng, 
 Minghui~Xu,
 Yue~Zhang, 
 Kun~Li, 
 Ruoxi~Wang, 
 Lian~Yang
 \thanks{Corresponding author: Minghui Xu.}
		\thanks{Y. Cheng, M. Xu, and K. Li are with the School of Computer Science and Technology, Shandong University, China (e-mail: \{yech@mail., mhxu@, kunli@\}sdu.edu.cn).}
		\thanks{Y. Zhang is with the Department of Computer Science, Drexel University, USA (e-mail: yz899@drexel.edu).}
        \thanks{R. Wang is with Northeastern University, USA (e-mail: wang.ruoxi2@northeastern.edu).}
        \thanks{L. Yang is with the Tendering and Procurement Department, Shandong Cancer Hospital and Institute, Shandong First Medical University and Shandong Academy of Medical Sciences, China (e-mail: wudimeishaonian37@126.com).}
	}
\IEEEtitleabstractindextext{%
\begin{abstract}

IoT platforms, particularly smart home platforms providing significant convenience to people's lives such as Apple HomeKit and Samsung SmartThings, allow users to create automation rules through trigger-action programming. However, some users may lack the necessary knowledge to formulate automation rules, thus preventing them from fully benefiting from the conveniences offered by smart home technology. To address this, smart home platforms provide pre-defined automation policies based on the smart home devices registered by the user. Nevertheless, these policies, being pre-generated and relatively simple, fail to adequately cover the diverse needs of users. Furthermore, conflicts may arise between automation rules, and integrating conflict detection into the IoT platform increases the burden on developers.
In this paper, we propose AutoIoT, an automated IoT platform based on Large Language Models (LLMs) and formal verification techniques, designed to achieve end-to-end automation through device information extraction, LLM-based rule generation, conflict detection, and avoidance. AutoIoT can help users generate conflict-free automation rules and assist developers in generating codes for conflict detection, thereby enhancing their experience. A code adapter has been designed to separate logical reasoning from the syntactic details of code generation, enabling LLMs to generate code for programming languages beyond their training data. Finally, we evaluated the performance of AutoIoT and presented a case study demonstrating how AutoIoT can integrate with existing IoT platforms.

\end{abstract}

}

\maketitle

\IEEEdisplaynontitleabstractindextext
\IEEEpeerreviewmaketitle

\section{Introduction} \label{sec:introduction}

With the development of embedded systems~\cite{pearson2020sic} and wireless communications~\cite{zhang2022good,zhang2020breaking}, IoT devices have been rapidly promoted and deployed worldwide~\cite{liu2024riotfuzzer,zhang2024collapse,lei2024friend}. The smart home has been enabled after manufacturers produced various kinds of smart IoT devices and developed IoT platforms. Mature IoT platforms include HomeKit~\cite{homekit}, SmartThings\cite{smartthings}, Mi Home\cite{mihome}, and Nest\cite{nest}. Users could use the platforms by setting up automation rules to fit their living habits, such as opening windows and playing music at 7:00 daily. The primary method of achieving automation rules is known as trigger-action programming. In trigger-action programming, users register and log into the IoT platform account, configure each device, and create automation rules through the user interface provided by the platform. Users do not need to understand the underlying code that the devices execute; instead, they only need to focus on the actions and states of the smart home devices to achieve automated control.  

Not all smart home users possess the necessary skills to create automation rules for IoT devices. Some users, due to a lack of knowledge or a disinclination about IoT devices, remain entirely unfamiliar with automation rules yet still desire to benefit from the conveniences that smart homes offer. 
Currently, mature IoT platforms (e.g., Mi Home) provide recommended automation rules based on the IoT devices registered by users. However, these rules are pre-defined relatively simply, so they may not meet users' needs.

Furthermore, the security problem exists among different automation rules. Consider the following example: one automation rule turns on the air conditioning when the temperature exceeds 25°C, which reduces the temperature and unexpectedly triggers another automation rule to activate the heater when the temperature drops below 25°C. There is a conflict between the two automation rules. Conflict detection is a significant issue in the smart home domain. The conflict detection approaches include formal modeling approach~\cite{9499364}~\cite{6963451}~\cite{IBRHIM2020106983}, graph-based approach~\cite{10.1145/3395363.3397347}~\cite{8975726}~\cite{9153388} and model-checking approach~\cite{wang2019charting}~\cite{DBLP:conf/ndss/CelikTM19}~\cite{10.1145/3368089.3409682}. However, the implementation of these conflict detection approaches requires manually writing relevant code for conflict detection, which increases the burden on developers.

The above research has significantly improved the usability and safety of smart home systems, but there are the following limitations: (1) Current IoT platforms require users to manually create automation rules, often lacking the capability to automatically generate rules that meet user needs. This is because the needs of different users vary significantly. Even though IoT platforms can assist users by providing predefined rules, these rules are often insufficient to meet the diverse requirements of individual users.
(2) The code for conflict detection often needs to be manually written, which is inefficient. The two limitations result in \textit{the current lack of a fully automated IoT platform that can achieve the entire process automation from rule generation, and conflict detection, to rule deployment}. Motivated by these pressing needs, in this paper, we present AutoIoT, an automated IoT platform using LLMs, that could generate and deploy conflict-free automation rules based on users' needs. Specifically, to achieve the generation of automation rules, we design an LLM-based rule generation component that significantly reduces the knowledge barrier for users to experiment with smart home automation. Users simply need to upload images of their deployed IoT devices to AutoIoT and then receive a series of automation rules. To ensure the safety of the rules, we also developed a conflict detection and avoidance component to make the rules conflict-free.

Our contributions are highlighted as follows: 
\begin{enumerate}
\item \textbf{End-to-end Automation.} We propose AutoIoT, the first one to achieve end-to-end automation through device information extraction, LLM-based rule generation, conflict detection, and avoidance. AutoIoT can be embedded seamlessly into existing IoT platforms and streamlines IoT management, saving time and effort for both users and developers.

\item \textbf{Conflict-free Rule Generation.} We present a novel LLM-based automation rule generation method. Our method rigorously formalizes four common rule conflict types, enabling proactive detection and resolution of potential issues. This ensures the generation of safe and reliable automation rules, eliminating the need for post-hoc conflict detection and correction.

\item \textbf{Developer-friendly Formal Verification.} For formal verification, we present a code generation adapter that separates logical reasoning from the syntactic details of code generation. This innovative approach enables LLMs to generate formal code for programming languages beyond their training data, freeing developers from the tedious task of writing intricate code for formal verification and significantly accelerating conflict detection.

\item \textbf{Implementation and Case Study.} We implement AutoIoT and evaluate it with different LLMs and combinations of IoT devices. We build a case study to demonstrate how AutoIoT is utilized for real-world smart home deployment, providing new directions for the future development of smart home systems.

\end{enumerate}

The rest of the paper is organized as follows. 
Section~\ref{sec:related work} provides a comprehensive review on the most relevant works. 
In Section~\ref{sec: Preliminaries}, we present the preliminary knowledge necessary to understand the subsequent sections.
Section~\ref{sec:LLMIOT} delves into details of the proposed AutoIoT, whose prototype implementation and performance evaluation are discussed in Section~\ref{sec:evaluation}.

\section{Related Works and Motivations} \label{sec:related work}
In this section, we first summarize the status quo of IoT Platforms in both academia and industry, then conduct an investigation on conflict detection technology for smart homes, and finally describe the motivations behind our design of an automated and safe smart home. 

\subsection{Mainstream IoT Platforms}

\begin{table*}[!th]
\centering
\caption{Comparison of Automated IoT Platform}\label{tab: relatework}
\begin{tabular}{cccccccc}
\toprule
\multirow{2}{*}{\textbf{IoT platforms}} & \multirow{2}{*}{\textbf{\begin{tabular}[c]{@{}c@{}}Platform \\ Compatibility\end{tabular}}} & \multicolumn{3}{c}{\textbf{Automation Rule}} & \multicolumn{3}{c}{\textbf{Conflict Resolution}} \\ \cline{3-8} 
 &  & \textbf{Functionality} & \textbf{Deployment Method} & \textbf{\begin{tabular}[c]{@{}c@{}}Automated \\ Generation\end{tabular}} & \textbf{Detection} & \textbf{Correction} & \textbf{\begin{tabular}[c]{@{}c@{}}Automated \\ Formal Verification\end{tabular}} \\
\midrule
 HomeKit~\cite{homekit} & \Checkmark   &  \Checkmark    &  APP / Script & \XSolidBrush & \XSolidBrush & \XSolidBrush & \XSolidBrush\\
\midrule
 SmartThings~\cite{smartthings} & \Checkmark   &  \Checkmark    &  APP / Script  & \XSolidBrush & \XSolidBrush & \XSolidBrush & \XSolidBrush\\
\midrule
 Mi Home~\cite{mihome} & \Checkmark   &  \Checkmark    &  APP / Script  & \XSolidBrush & \XSolidBrush & \XSolidBrush & \XSolidBrush\\
 \midrule
 Nest~\cite{nest} & \Checkmark   &  \Checkmark    &  APP / Script  & \XSolidBrush & \XSolidBrush & \XSolidBrush & \XSolidBrush\\
 \midrule
 Alexa~\cite{amazon} & \Checkmark   &  \Checkmark    &  APP / Script  & \XSolidBrush & \XSolidBrush & \XSolidBrush & \XSolidBrush\\
 \midrule
 Framework~\cite{esposito2023design} & \Checkmark  &  \Checkmark    & Script  & \XSolidBrush & \XSolidBrush & \XSolidBrush & \XSolidBrush\\
 \midrule
 IoTMediator~\cite{287119} & \Checkmark   &  \Checkmark    &  APP / Script  & \XSolidBrush & \Checkmark  & \Checkmark & \XSolidBrush\\
 \midrule
 CP-IoT~\cite{lincp} & \Checkmark   &  \Checkmark     &  APP  & \XSolidBrush & \Checkmark & \Checkmark & \XSolidBrush\\
 \midrule
 AutoIoT (Ours) & \Checkmark   &  \Checkmark    &  APP / Script  & \Checkmark & \Checkmark  & \Checkmark & \Checkmark\\
\bottomrule
\end{tabular}
\end{table*}

To clarify the capabilities of mainstream IoT platforms, as shown in Table~\ref{tab: relatework}, we evaluate them based on the following seven aspects: compatibility with commercial platforms, support for automation rules, deployment methods for rules, automatic generation of rules, conflict detection for rules, conflict resolution and automatic generation of conflict detection code.

The development of IoT Platforms in the industry primarily focuses on enhancing the intelligence of smart home device control to provide seamless, efficient, and user-friendly experiences. HomeKit\cite{homekit} integrates seamlessly with Apple devices, including iPhones, iPads, and Macs, providing a cohesive user experience. SmartThings\cite{smartthings} supports multiple communication protocols, including Zigbee, Z-Wave, and Bluetooth, ensuring compatibility with a broad range of devices. Philips Hue\cite{philips} is known for its advanced smart lighting solutions, offering a wide range of bulbs, strips, and fixtures that can be controlled via the Hue app or voice assistants. Amazon Alexa\cite{amazon}, Mi Home\cite{mihome}, and Nest\cite{nest} could allow users to control smart home devices using simple voice commands and automatically adjust some functions, such as temperature settings, to save energy and improve comfort.

In academia, researchers focus on different aspects of IoT platforms such as architecture and security. They also conduct experiments to evaluate scenario usability and propose solutions to address various challenges. Esposito~\textit{et al.}~\cite{esposito2023design} proposed a framework for smart home automation based on MQTT and cloud-deployed serverless functions. Vellela~\textit{et al.}~\cite{10270407} presented a cloud-based, intelligent IoT platform for a customized healthcare data gathering and monitoring system. Chi~\textit{et al.}~\cite{287119} presented IoTMediator, which provides accurate threat detection and threat-tailored handling in multi-platform multi-control-channel homes. Liu~\textit{et al.}~\cite{10255286} proposed a Tokoin-based access control model for IoT enabled by blockchain. Wang~\textit{et al.}~\cite{10228993} presented IoTDuet for determining where an IoT device event is triggered to achieve smart home safety monitoring. Lin~\textit{et al.}~\cite{lincp} designed CP-IoT to monitor the execution behavior of the automation and discover the anomalies, as well as hidden risks among them on heterogeneous IoT platforms.

\subsection{Conflict Detection for Smart Home}
Conflict detection in the smart home is a critical process aimed at identifying and resolving situations where the commands or settings of multiple devices or services within the smart home environment are incompatible or contradictory. These conflicts can arise due to various factors, such as conflicting automation policies, competing device settings, or conflicting user preferences. Effective conflict detection ensures that the smart home operates smoothly and safely, preventing issues that could lead to user frustration, inefficiency, or even potential hazards.

Ding~\textit{et al.}~\cite{ding2021iotsafe} proposed a dynamic safety and security policy enforcement system IoTSAFE to capture and manage real physical interactions on smart home platforms. Celik~\textit{et al.}~\cite{DBLP:conf/ndss/CelikTM19} presented IoTGUARD to dynamically detect unsafe and insecure interactions between IoT apps. Wang~\textit{et al.}~\cite{wang2019charting} designed iRuler, which performs satisfiability modulo theories solving and model checking to discover inter-rule vulnerabilities within IoT deployments. Xiao~\textit{et al.}~\cite{9800202} proposed a digital-twin-based preventive framework to defend against the threats caused by malicious apps in smart home systems. Huang~\textit{et al.}~\cite{huang2023survey} discussed the conflict detection approaches and open issues on IoT-based smart homes. Fang~\textit{et al.}~\cite{FANG2021100004} presented a model checking-based framework to comprehensively evaluate IoT system security.

\subsection{Limitations}
Based on the above summary and analysis, one can see that mainstream IoT platforms have enhanced user experience through various interaction methods such as voice control and mobile app control. However, these platforms do not support the automatic generation of rules based on users' needs, nor do they support automatic conflict detection of automation rules. This increases the user threshold for smart home devices to some extent, limiting the ability of individuals who lack knowledge about smart home technology, such as some older adults, to benefit from the conveniences these devices offer, thereby affecting the promotion and widespread adoption of smart homes. Additionally, current research related to conflict detection often requires developers to write specific detection codes or train relevant models, which cannot be seamlessly integrated with existing IoT platforms.

\section{Preliminaries}\label{sec: Preliminaries}
In this section, we present the preliminary of LLM, prompt engineering, and formal verification, which are the basic building blocks of our work.

\subsection{LLM and Prompt Engineering}
LLMs~\cite{zhao2024surveylargelanguagemodels,li2024attention} represent a significant leap forward in the capabilities of language models, driven by advances in deep learning and the availability of large-scale training data. LLMs are characterized by several distinctive features:
\begin{enumerate}
\item \textbf{Massive Scale}: LLMs typically encompass tens or hundreds of billions of parameters, far surpassing the scale of earlier models like BERT. This substantial increase in model size allows LLMs to capture more nuanced and complex patterns within large datasets, thereby enhancing their capacity for language understanding and generation.
\item \textbf{Emergent Abilities}: One of the most intriguing aspects of LLMs is their ability to exhibit emergent behaviors that were not explicitly programmed or anticipated. 
\item \textbf{Contextual Understanding}: LLMs excel at understanding context within natural language. They can infer meaning from complex sentences, understand relationships between entities, and maintain coherence over long stretches of text. This contextual understanding is crucial for tasks such as open-domain question answering, document summarization, and dialogue systems.
\item \textbf{Advanced Reasoning Capabilities}: LLMs are capable of performing sophisticated reasoning tasks, including logical deduction, mathematical problem-solving, and even creative writing.
\end{enumerate}
Such capabilities have driven notable progress in Natural Language Processing (NLP)~\cite{feng2021surveydataaugmentationapproaches}, leading to diverse applications across sectors including programming~\cite{cai2024lowcodellmgraphicaluser}, code understanding~\cite{10.1145/3597503.3639187}, and 3D physical world modeling~\cite{hong20233d} and medicine~\cite{alberts2023large}. Additionally, LLMs also propel advancements in research related to security and privacy~\cite{yan2024protectingdataprivacylarge}~\cite{YAO2024100211,wang2024conu,wang2024word}.

Prompt engineering is a specialized discipline within the field of artificial intelligence and machine learning, focusing on the design and optimization of prompts to elicit desired responses from AI models, particularly LLMs. In this context, a prompt refers to the input text provided to an AI model to generate a specific output or perform a particular task. The elements of a prompt \cite{aiprompt} include the following:
\begin{enumerate}
\item \textbf{Instruction} is a specific task or directive that you want the AI model to perform. It clearly communicates the goal or objective of the interaction with the model.
\item \textbf{Context} refers to external information or additional background that can help the model provide more accurate and relevant responses.
\item \textbf{Input data} is the specific input or question that you are interested in getting a response for. It is the core content that the model will process to generate an output.
\item \textbf{Output indicator} specifies the type or format of the output you expect from the model. It guides the model on how to structure its response. 
\end{enumerate}

\begin{figure*}[!htbp]
    \centering
    \centerline{\includegraphics[width=1.0\textwidth, ]{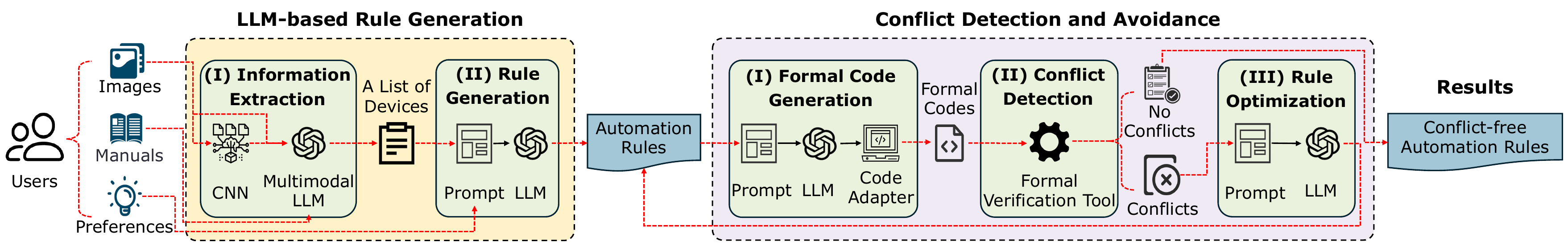}}
	\caption{The architecture of AutoIoT.} 
	\label{fig: architecture}
\end{figure*}

Understanding these elements is crucial because they allow us to effectively communicate our intentions to the model. By carefully crafting the prompt, we can guide the model’s behavior and improve the quality of its responses. 
These elements provide the necessary structure and context for the model to generate accurate and meaningful outputs\cite{giray2023prompt}.

\subsection{Formal Verification}
Formal verification in smart home systems refers to the process of using mathematical techniques to prove or disprove the correctness of algorithms, software, and hardware systems that are part of a smart home system. The goal of formal verification is to ensure that these systems operate as intended without errors or vulnerabilities that could lead to security breaches, malfunctions, or safety issues. The process of formal verification typically involves creating a precise mathematical model of the system and then using formal methods—such as model checking, theorem proving, and static analysis—to analyze this model against a set of specifications or requirements.

Maude \cite{Maude} is a high-performance reflective language and system designed to support both equational and rewriting logic specification and programming across a broad spectrum of applications. Rooted in the legacy of the OBJ3 language, which itself can be seen as an equational logic sublanguage, Maude extends beyond its predecessor by integrating support for rewriting logic computation, thereby enriching its capability to model and reason about concurrent and non-deterministic systems \cite{wang2019charting}.

\section{AutoIoT} \label{sec:LLMIOT}
\subsection{Modeling Devices and Rules}  \label{sec:model}
In smart homes, users interact with devices to enjoy convenience. A smart home platform serves as an intermediary, connecting users to devices and enabling control through various methods, including smartphone apps, manual interaction, voice commands, and automation. This paper focuses on the automation aspect. Smart home devices are categorized into two types: sensors, which measure environmental conditions, and actuators, which execute user commands. To formally describe device combinations for automation, we introduce a device model and define an automation rule.
A smart home device is formally defined as a tuple $$\mathsf{Device}=(\mathsf{ID}, \mathsf{Type}, \mathsf{Action}, \mathsf{State}, \mathsf{Location}).$$ Here, $\mathsf{ID}$ uniquely identifies the device, $\mathsf{Type}$ is the category of the device (e.g., light, thermostat, sensor), $\mathsf{Action}=\{a_1, a_2,...\}$ is the set of actions the device can perform, $\mathsf{State}=\{s_1, s_2,...\}$ is the set of possible states the device can be in, and $\mathsf{Location}$ refers to the room or area where the device is located. For instance, a smart light bulb might have actions like ``turn\_on'', ``turn\_off'', ``dim'', and ``brighten'', and states like ``off'', ``on'', ``low'', and ``high''. This formalization provides a precise and structured representation of smart home devices, facilitating their modeling, control, conflict detection, and automation.

Automation rules of a smart home system can be represented as a set $\{R_i\}$, where $R_i$ denotes the $i$-th rule. An automation rule $R_i$ is defined as follows:
$$ R_i: \ \mathbf{If} \ T(S_i, E_i) = 1, \ \mathbf{then} \ S_i \stackrel{A_i}{\longrightarrow} S_i',$$ where $T$ is a trigger function that determines whether the initial states $S_i$ and environmental conditions $E_i$ satisfy the rule's trigger conditions. In specific, $S_i=\{s_1, s_2,...\}$ is a set of initial states of relevant devices, and $E_i=\{e_1, e_2,...\}$ is a set of environmental factors, such as temperature, humidity, light intensity, or time, measured by sensors. $A_i=\{a_1, a_2,...\}$ is a set of actions to be executed. The resulting states of the devices after executing the actions are denoted by $S_i'=\{s_1', s_2',...\}$. For example, an automation rule might be: ``If the air conditioner is off and the temperature is above 27 degrees Celsius, then execute the action \texttt{turn\_on} to turn on the air conditioner."

Our objective is to develop techniques to generate conflict-free automation rules using LLMs. The attacker might compromise LLMs, where the training data may have been subjected to poisoning or backdoor attacks, thereby preventing the model from generating conflict-free automation rules upon prompt. However, we will not delve into how such LLMs become compromised, as it is out of the scope of this paper and there are also LLM vendors who can adopt various countermeasures to deal with such attacks. 

\subsection{The Overview of AutoIoT}
Based on our findings and the rapid advancement of LLMs, we recognize that LLMs excel in understanding complex logic and reasoning, making them highly suitable for generating automation policies on behalf of users and detecting conflicts between policies. Therefore, we propose AutoIoT, which achieves full automation including device information extraction, rule generation, code generation, conflict detection, and conflict avoidance. Users only need to provide images of IoT devices to obtain conflict-free automation rules that meet their needs, significantly lowering the knowledge barrier for users and developers. As shown in Fig. \ref{fig: architecture}, AutoIoT consists of two components:

\begin{enumerate}
\item \textbf{LLM-based Rule Generation (\ref{sec:LRG}).} 
This component takes diverse inputs, including images of IoT devices (app screenshots or user-captured photos), device manuals, and user-specified automation preferences. Using a combination of CNN models and LLMs, it first identifies the types, locations, actions, and states of the IoT devices. This information is then structured into a JSON list, which, along with the user's preferences, serves as a prompt for the LLM. The LLM processes this prompt to generate a list of potential automation rules, tailored to the user's specific needs.
\item \textbf{Conflict Detection and Avoidance (\ref{sec:CDA}).} 
This component takes a set of automation rules as input and generates a refined set of safe automation rules. It leverages formal verification techniques to analyze and optimize the input rules. First, the rules are translated into formal specifications, which are then formally verified to identify potential conflicts. Subsequently, an LLM is employed to resolve these conflicts, generating a new set of safe and optimized automation rules. This ensures that the generated rules are not only correct but also harmonious with the overall system's behavior, namely conflict-free.
\end{enumerate}

\subsection{LLM-based Rule Generation} \label{sec:LRG}
To generate automation rules, AutoIoT employs a two-step process. First, it extracts $\langle \mathbf{device\_list} \rangle$,  a list of IoT devices from user-provided images (screenshots or photos) and device manuals. Next, it leverages these device details, combined with user preferences, to construct prompts for an LLM. The LLM, in turn, generates $\langle \mathbf{rule\_list} \rangle$, consisting of tailored automation rules that align with the user's specific needs and device capabilities.

\subsubsection{Device Information Extraction} \label{sec:IR}

\begin{figure}[!htbp]
    \subfloat[]{\includegraphics[width=0.132\textwidth]{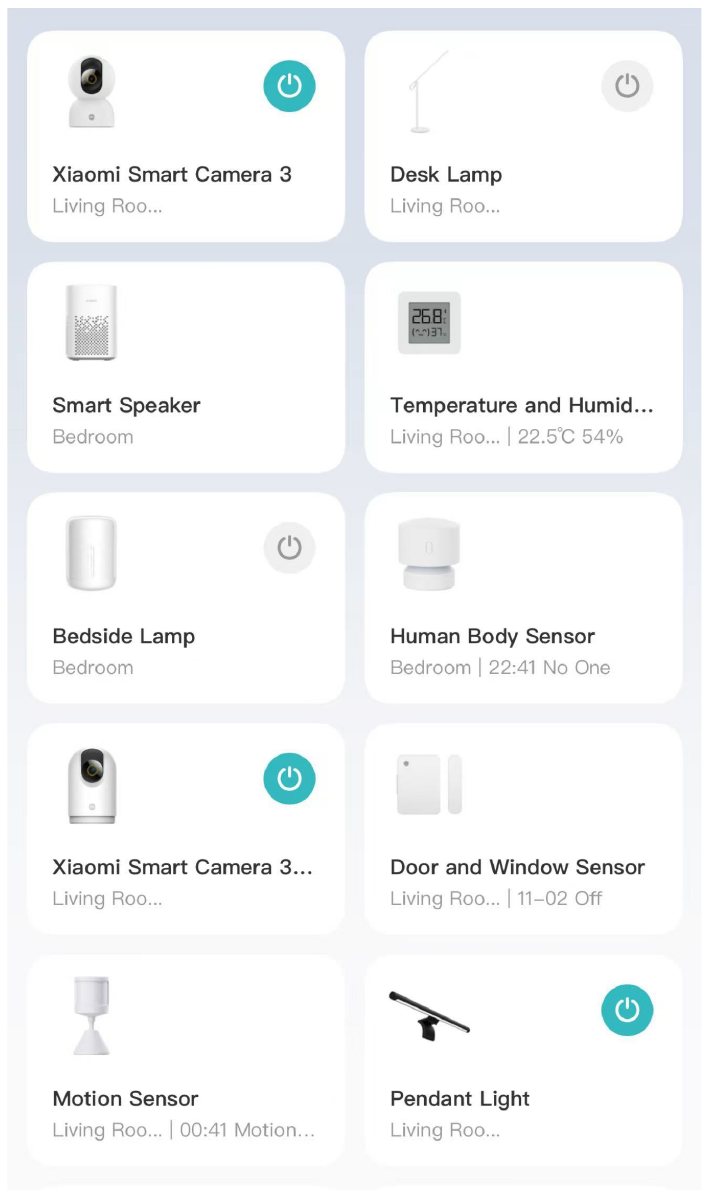}\label{fig:images:screenshot}}
    \subfloat[]{\includegraphics[width=0.35\textwidth]{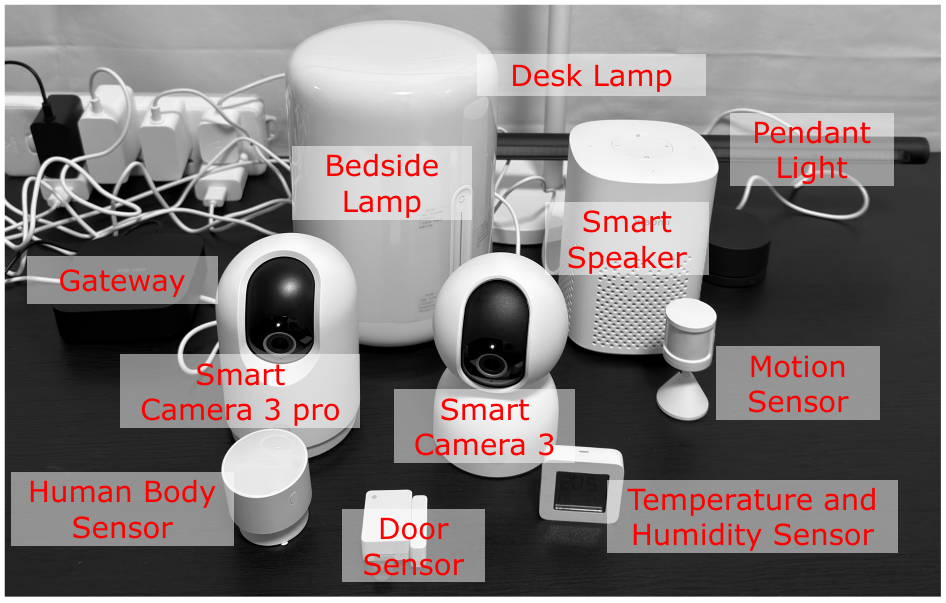}\label{fig:images:photo}}
    \caption{Two types of images for device information extraction. (a) App screenshots. (b) User-captured photos.} 
    \label{fig:image}
\end{figure}
AutoIoT first extracts information from uploaded images, including IDs, types, actions, states, and locations of the user's IoT devices. Based on whether the IoT platform provides a graphical user interface to control the IoT devices, we categorize the images users upload into two types. The first type of image comprises screenshots from smart home app interfaces as shown in Fig.~\ref{fig:images:screenshot}, typically containing information like IDs, types, and location information of each device. In some cases, where informative app interfaces are unavailable, we can directly extract device information from actual photos taken by users, as depicted in Fig.~\ref{fig:images:photo}. 

\begin{table}[!htbp]  
\centering  
\caption{Prompt $\mathcal{A}$: Device Information Extraction}\label{tab:list_gen}  
\begin{tabular}{p{0.2\columnwidth}p{0.7\columnwidth}} 
\toprule  
\textbf{Elements} & \textbf{Contents} \\  
\midrule  
\textbf{Instruction} &
Based on the uploaded manuals, identify the states and actions of each IoT device in the input data according to Context.\\ 
\midrule  
\textbf{Context} &
Backgrounds:

A smart home device is formally defined as a tuple Device = (ID, Type, Action, State, Location). Here, ID uniquely identifies the device, Type is the type of the IoT device, Action is the set of operations it can perform, State is the set of its possible states, and Location refers to the room in which the device is deployed. 

For instance, a smart light bulb might have operations like ``turn on'', ``turn off'', ``dim'', and ``brighten'', and states like ``off'', ``on'', ``low'', and ``high''.\\

\midrule  
\textbf{Input data} & $\langle \mathbf{incomplete\_device\_list} \rangle$, $\langle \mathbf{user\_manual} \rangle$.\\  
\midrule  
\textbf{Output indicator} &   $\langle \mathbf{device\_list} \rangle$ in the JSON format:

\{``ID1'':\{``type'':``device type'', ``action'':[``a1'', ``a2'', ...], ``state'':[``s1'', ``s2'', ...], ``location'':``room1''\}, ``ID2'':\{...\}, ...\}. \\  
\bottomrule  
\end{tabular}  
\end{table}

\begin{table*}[!t]  
\centering  
\caption{Prompt $\mathcal{B}$: Automation Rule Generation}\label{tab:rule_gen}  
\begin{tabular}{p{0.13\textwidth}p{0.8\textwidth}} 
\toprule  
\textbf{Elements} & \textbf{Contents} \\  
\midrule  
\textbf{Instruction} & Generate conflict-free automation rules based on the Input data according to Context, while considering the user's preferences.\\ 
\midrule  
\textbf{Context} & 
Basic backgrounds:

1. A smart home device is formally defined as a tuple Device = (ID, Type, Action, State, Location). Here, ID uniquely identifies the device, Type is the type of the IoT device, Action is the set of operations it can perform, State is the set of its possible states, and Location refers to the room in which the device is deployed. For instance, a smart light bulb might have operations like ``turn on'', ``turn off'', ``dim'', and ``brighten'', and states like ``off'', ``on'', ``low'', and ``high''.

2. An automation rule is represented as (triggers, actions). A trigger can be a device state, an environmental condition (e.g., temperature, brightness, humidity, or sound), a specific time, or a particular location, etc. Actions are the operations that the device can execute. 

Additional backgrounds:

3. Conflicts consist of state conflict (two rules attempt
to change the state of the same device in incompatible ways), environment conflict (two rules have conflicting effects on the environment), state cascading conflict (the resulting states of one rule become the initial states for
another rule), and state-environment cascading conflict (some rules’ resulting states and environment factors can trigger another rule).\\  
\midrule  
\textbf{Input data} &  $\langle \mathbf{device\_list} \rangle$, $\langle \mathbf{user\_preference} \rangle$.\\  
\midrule  
\textbf{Output indicator} &  $\langle \mathbf{rule\_list} \rangle$ in the JSON format:\{``rule1'': \{``trigger'': ``xxx'', ``action'': ``xxx''\} ... \}. \\  
\bottomrule  
\end{tabular}  
\end{table*} 

To identify device IDs, types, and locations from screenshots, we employ a multimodal LLM to extract textual information from the screenshots. This approach generates $\langle \mathbf{incomplete\_device\_list} \rangle$, a list of device IDs and their corresponding types and locations. We choose multimodal LLM instead of optical character recognition (OCR) because the former enables deeper context and comprehension and thus can output complete and structured information based on multiple screenshots. 
For user-captured photos, where LLMs are less effective for image recognition, we train a Convolutional Neural Network (CNN) model, specifically a ResNet-18 architecture, to perform image classification. We create a dataset consisting of photos categorized into 10 classes, including sensors, lamps, speakers, and cameras. The CNN model takes a photo as input and outputs $\langle \mathbf{incomplete\_device\_list} \rangle$, a list of predicted device types. 
To further extract the actions and states, we formulate a prompt designed for LLM interaction, as outlined in Table~\ref{tab:list_gen}. The prompt instructs the LLM to identify device states and actions based on the uploaded device manuals. The prompt's context contains device definitions. The expected output $\langle \mathbf{device\_list} \rangle$ is a list of devices in the JSON format.

\subsubsection{Rule Generation}
Next, AutoIoT uses a list of IoT devices and user-specified automation preferences (e.g., avoiding turning on the light at midnight) as inputs and outputs a list of automation rules. 
Considering the text understanding and reasoning capabilities of LLMs, we utilize LLMs to generate automation rules that cater to different user preferences. Moreover, LLMs have the ability to infer specific automation rules from abstract user preferences. For example, a user might simply state a desire for the home environment to be more comfortable in the morning. LLMs could interpret this vague requirement and, taking into account factors such as the current season and weather conditions, generate a series of rules, such as adjusting the room temperature by turning on the air conditioner and turning on soft lighting.

To guide the LLM in generating automation rules, we design a prompt (Table~\ref{tab:rule_gen}) that provides specific instructions. The prompt includes background information on device definitions and automation rule formats to ensure accurate understanding. Additional background information for conflict resolution will be discussed in Section~\ref{sec:CDA}. The input data consists of a device list ($\langle \mathbf{device\_list} \rangle$) and a user preference list ($\langle \mathbf{user\_preference} \rangle$). If the user has no specific preferences, $\langle \mathbf{user\_preference} \rangle$ can be omitted. The output, $\langle \mathbf{rule\_list} \rangle$, is expected to be a list of automation rules in JSON format.

\begin{figure}[!htbp]
    \centering
    \centerline{\includegraphics[width=0.48\textwidth]{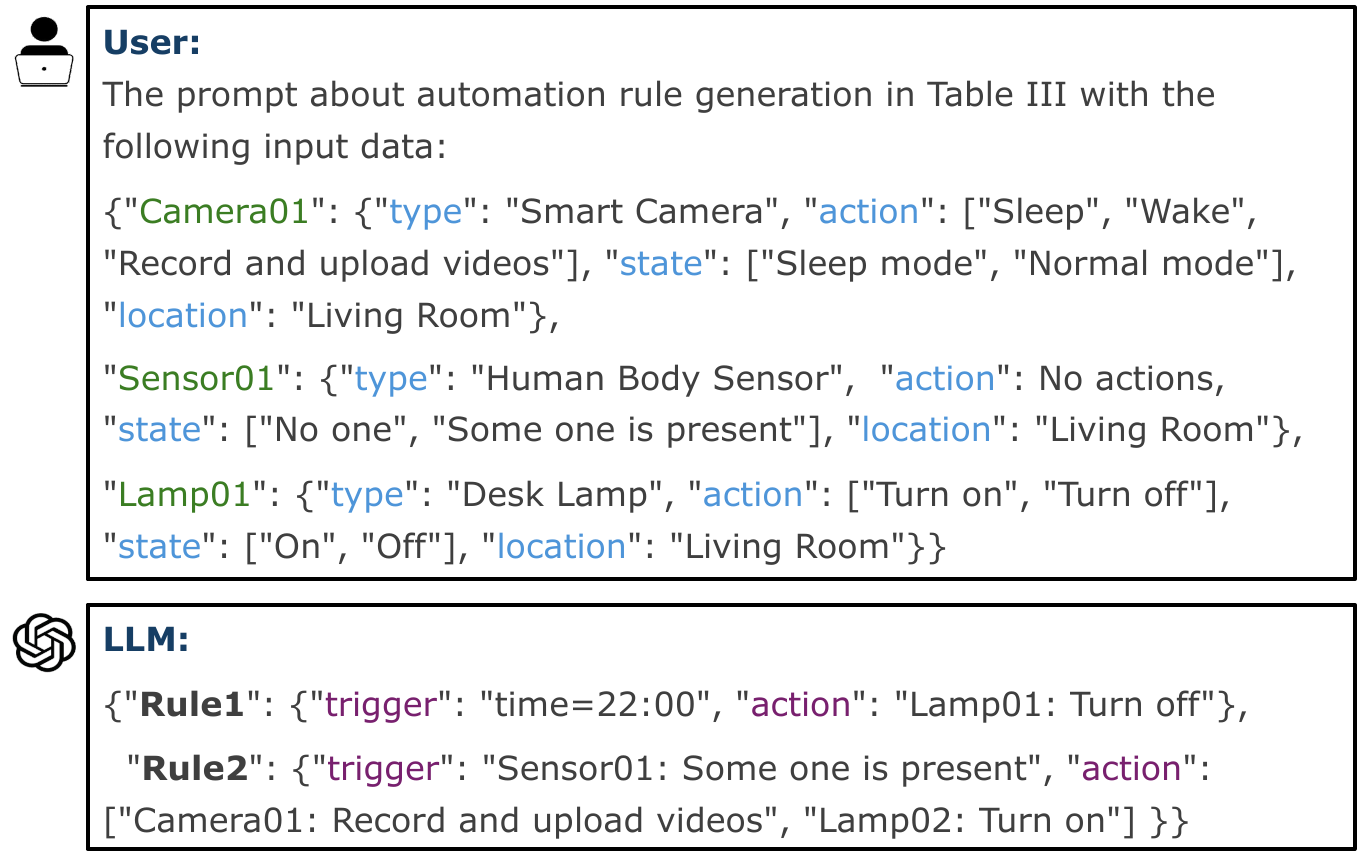}}
	\caption{An example of LLM-generated automation rules.} 
	\label{fig:prompt2}
\end{figure}

We interact with the LLM using this prompt. As illustrated in Fig.~\ref{fig:prompt2}, after providing a specific use case as input, the LLM generates a list of automation rules in the required format. AutoIoT supports users to configure IoT devices using natural language, significantly reducing the knowledge barrier and improving user experience. Besides, when users need to add new IoT devices or change their preferences, they can simply upload a new photo or send a new preference to AutoIoT. This allows for automatic adjustment of the automation rules without the need for manual reconfiguration.

\subsection{Conflict Detection and Avoidance}\label{sec:CDA}
Conflicts may arise among the rules in the $\langle \mathbf{rule\_list} \rangle$. For instance, in Fig.~\ref{fig:prompt2}, $\mathbf{Rule1}$ turns off the Desk Lamp in the living room at 22:00, potentially dimming the room's overall lighting. However, $\mathbf{Rule2}$ activates the smart camera in the living room to start recording when the bedroom sensor detects motion, requiring sufficient light. While $\mathbf{Rule1}$ and $\mathbf{Rule2}$ don't directly conflict in terms of timing or the devices they control, they can indirectly conflict by affecting environmental factors like lighting conditions. It is very common to encounter conflicts like this, and even more complex ones, in smart home setups.

To make the automation rules conflict-free, considering the strong semantic understanding and logical reasoning capabilities of LLMs, we first need to define the conflicts and add the definitions to the prompt about rule generation to guide LLMs in generating conflict-free automation rules. However, we cannot guarantee that the generated rules are entirely free from conflicts. Therefore, we next need to detect conflicts among these rules. Since executing automation rules before conflict detection is not a prudent approach, we opt for formal verification to detect conflicts. Furthermore, generating code is also a strong capability of LLMs. Utilizing LLMs to generate formal verification codes can significantly enhance the efficiency of conflict detection. Finally, we use the results of the conflict detection to determine whether rule optimization is necessary. This is a three-step process, starting from formal code generation, followed by conflict detection using formal verification, and finally conflict avoidance through prompt tuning.

\begin{figure*}[!htbp]
\centering
    \centerline{\includegraphics[width=0.95\textwidth]{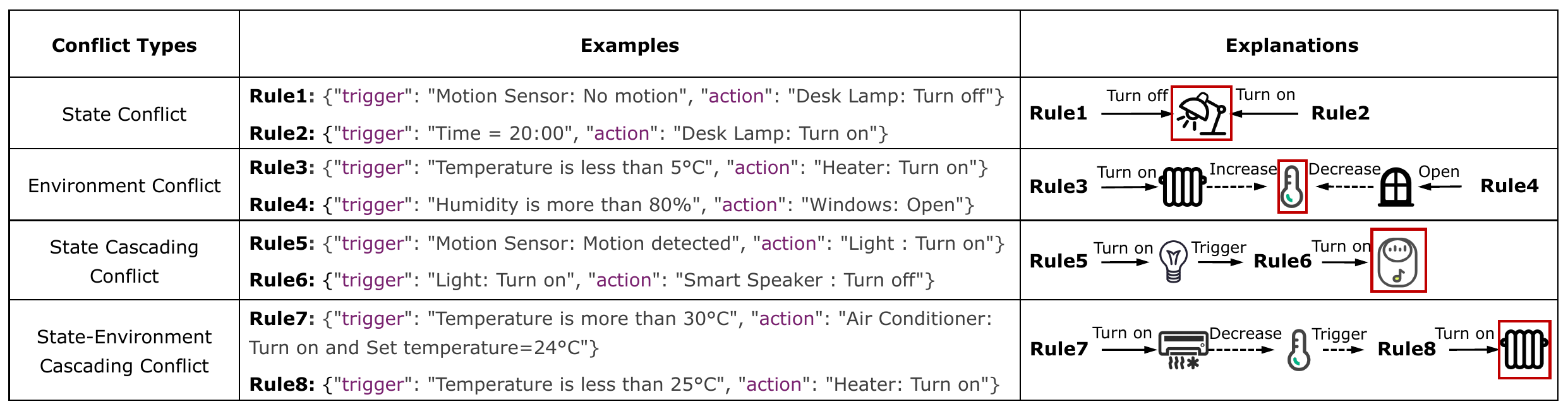}}
	\caption{The examples of four different conflicts.} 
	\label{fig:conflict}
\end{figure*}

\subsubsection{LLM-based Formal Code Generation}
To produce formal codes for conflict detection, we first provide a clear definition of four types of conflicts consisting of state conflict, environment conflict, state cascading conflict, and state-environment cascading conflict. We formalize conflicts based on our model and align our definitions with well-accepted ones \cite{lincp, ding2021iotsafe, huang2023survey}.

\begin{enumerate}
    \item \textbf{State Conflict.} For two automation rules $R_i$ and $R_j$, a potential state conflict may arise when both rules attempt to change the state of the same device in incompatible ways. For example, one rule might try to turn a light on, while another tries to turn it off. We use $C_{S}$ to denote a set containing all pairs of conflicting states for all devices. The set can be obtained through an exhaustive enumeration, where each element $(s_k,s_l)$, represents a pair of conflicting states. Formally, this conflict can be expressed as $\exists s_i', s_j' \in S_i' \cup S_j', (s_i', s_j') \in C_{S}$. For example, for a light bulb with four states: ``on'', ``off'', ``low'', and ``high'', the pairs (``on'', ``off'') and (``low'', ``high'') would belong to the set $C_{S}$. 

    \item \textbf{Environment Conflict.} For two automation rules $R_i$ and $R_j$, an environment conflict occurs when the two rules have conflicting effects on the environment. For instance, one rule might increase the temperature, while another decreases it simultaneously. $C_{E}$ denotes a set containing all pairs of conflicting states for all devices related to environmental factors. Formally, this conflict can be expressed as $\exists s_i', s_j' \in S_i' \cup S_j', (s_i', s_j') \in C_{E}$. 
    
    \item \textbf{State Cascading Conflict.} For automation rules $R_1$, $R_2$, ..., $R_k$, a state impact conflict arises when the resulting states of one rule become the initial states for another rule. This interdependency can lead to unintended cascading effects, where the execution of one rule triggers the execution of others. This can result in unexpected behaviors, potential resource exhaustion, or even system instability. Formally, a state Cascading conflict exists if $\forall s_i \in S_i, s_i \in S_1' \cup S_2' \cup \cdots \cup S_k'$.

    \item \textbf{State-Environment Cascading Conflict.} For automation rules $R_1$, $R_2$, ..., $R_k$, a state-environment cascading conflict occurs when some rules' resulting states and environment factors can trigger another rule. Formally, a conflict exists if $T(S_i, E_i)=1$ and $\forall s_i \in S_i, s_i \in S_1' \cup S_2' \cup \cdots \cup S_k'$.

\end{enumerate}

Based on the aforementioned conflicts, to improve the efficiency of conflict detection, we modify the prompt about rule generation to minimize conflicts during the generation of automation rules. As indicated in Table \ref{tab:rule_gen}, we add the definitions of the conflicts as additional backgrounds to the context. Thus, the context of the new prompt includes basic backgrounds and additional backgrounds. The instruction, the input data, and the output indicator are the same as the original ones, respectively. After generating the conflict-free automation rules, to verify that there indeed are no conflicts, we employ formal verification methods to detect potential conflicts among the rules.

\begin{table*}[!htbp]  
\centering  
\caption{Prompt $\mathcal{C}$: Formal Code Generation}\label{tab:code_gen}  
\begin{tabular}{p{0.14\textwidth}p{0.8\textwidth}} 
\toprule  
\textbf{Elements} & \textbf{Contents} \\  
\midrule  
\textbf{Instruction} &
Generate formal codes to detect conflicts among the automation rules based on the Input data according to Context.
\\ 
\midrule  
\textbf{Context} & 
Basic backgrounds:

1. A smart home device is formally defined as a tuple Device = (ID, Type, Action, State, Location). Here, ID uniquely identifies the device, Type is the type of the IoT device, Action is the set of operations it can perform, State is the set of its possible states, and Location refers to the room in which the device is deployed. For instance, a smart light bulb might have operations like ``turn on'', ``turn off'', ``dim'', and ``brighten'', and states like ``off'', ``on'', ``low'', and ``high''.

2. Conflicts consist of state conflict (two rules attempt
to change the state of the same device in incompatible ways), environment conflict (two rules have conflicting effects on the environment), state cascading conflict (the resulting states of one rule become the initial states for
another rule), and state-environment cascading conflict (some rules’ resulting states and environment factors can trigger another rule).

Additional backgrounds:

3. $\langle \mathbf{code\_template} \rangle$\\  
\midrule  
\textbf{Input data} & 
$\langle \mathbf{device\_list} \rangle$, $\langle \mathbf{rule\_list} \rangle$.\\  
\midrule 
\textbf{Output indicator} & 
$\langle \mathbf{logic\_code} \rangle$ in the format of code generation template.
\\  
\bottomrule  
\end{tabular}  
\end{table*} 

After conducting a thorough investigation, we discover that Maude~\cite{Maude} shares similarities with our formal representation of automation rules. Maude is a high-performance system for rewriting logic and functional programming, which makes it well-suited for formal verification tasks. By leveraging Maude's capabilities, we can effectively model and verify the safety of our automation rules.
Next, to improve the efficiency of producing codes, we write a prompt, to guide LLM to generate the Maude codes. We first define the instruction as generating Maude codes to detect conflicts among the automation rules based on the input data according to the context. The input data consists of a list of devices $\langle \mathbf{device\_list} \rangle$ and a list of automation rules $\langle \mathbf{rule\_list} \rangle$. The output is the codes that conform to Maude's syntax for formal verification. To make LLM understand the task clearly, we add our definition of a smart home device and information about Maude to the context. We interact with LLM using the above prompt. However, the generated codes contain numerous syntactical errors, such as ``undeclared sort'', ``bad token'', and ``no parse for statement''. We conjecture the cause is the lack of Maude-related data in LLM's training dataset. Additionally, it is challenging to find Maude projects in all available open-source code, making it difficult to construct a fine-tuning dataset.

\begin{figure}[!htbp]
    \centering
    \centerline{\includegraphics[width=0.48\textwidth]{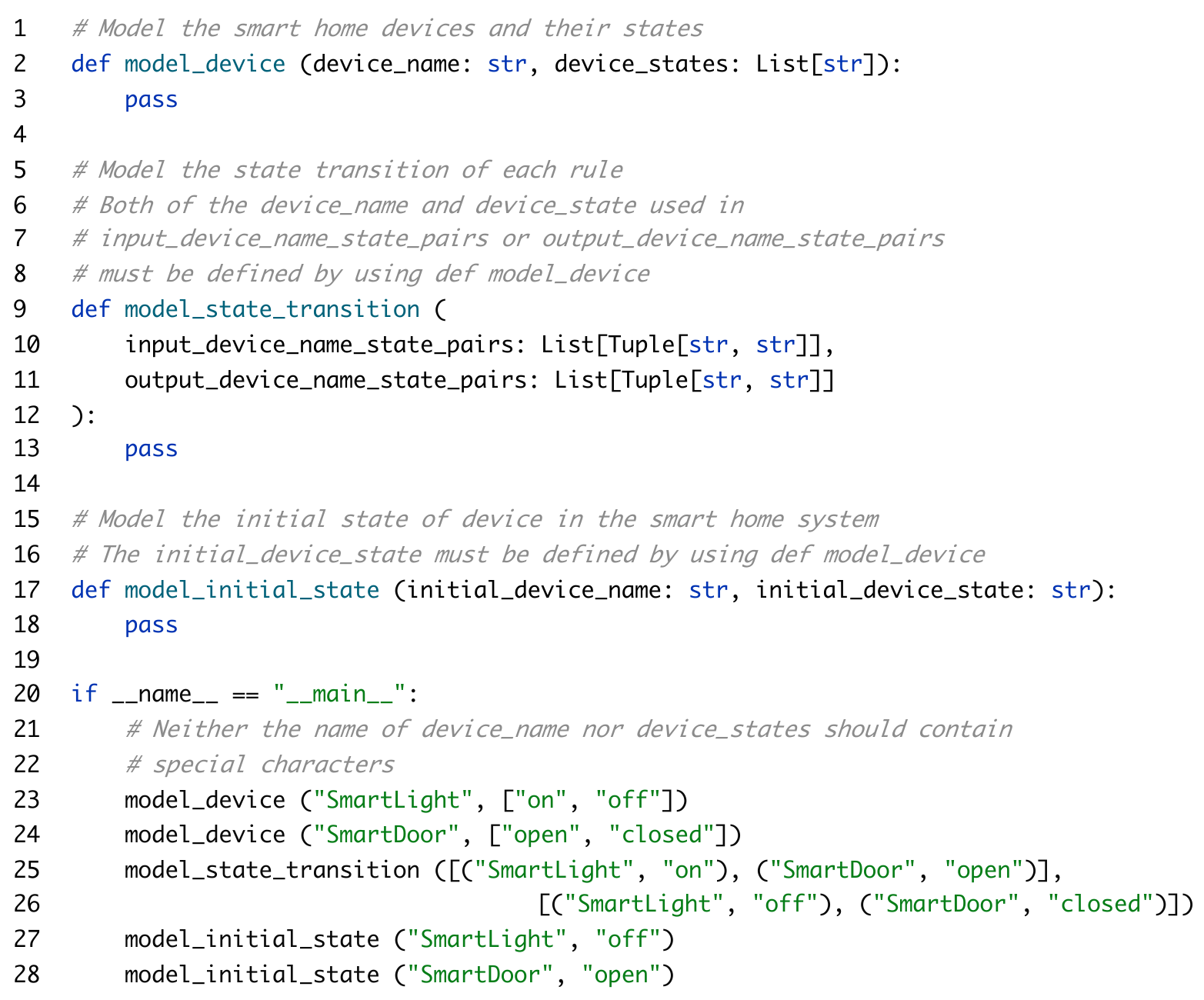}}
	\caption{The code generation template for formal verification.} 
	\label{fig:template}
\end{figure}

Based on these findings, we propose a code generation adapter to separate logic from syntax, ensuring that AutoIoT can generate Maude code that is both logically and syntactically correct. The details are described below.
To ensure that LLM can more easily and accurately understand the logic of Maude code generation, we abstract the logic into a formal code generation template, as shown in Fig.~\ref{fig:template}, which consists of three Python functions, designated as \texttt{model\_device}, \texttt{model\_state\_transition}, and \texttt{define\_initial\_state}. 
The function \texttt{model\_device} is responsible for modeling a device through its ID and states. 
The function \texttt{model\_state\_transition} models state transitions, requiring as input the initial and final states of all devices involved in the transition process upon invocation. 
The function \texttt{define\_initial\_state} defines the initial state of the entire smart home system.

Then, as indicated in Table~\ref{tab:code_gen}, we formulate a prompt designed for formal code generation. The prompt instructs the LLM to generate formal codes based on $\langle \mathbf{device\_list} \rangle$ and $\langle \mathbf{rule\_list} \rangle$. We add device definitions, conflict definitions as basic backgrounds, and the code generation template $\langle \mathbf{code\_template} \rangle$ as additional backgrounds to the prompt's context. The expected output $\langle \mathbf{logic\_code}\rangle$ is Python codes in the format of $\langle \mathbf{code\_template} \rangle$. With the prompt, LLM only needs to generate the formal verification logic by calling the functions in the code generation template, without caring about the specific code syntax of Maude. 
After receiving the codes produced by LLM in the format of the code generation template, AutoIoT uses the code generation adapter to convert $\langle \mathbf{logic\_code} \rangle$ into Maude codes for conflict detection. The details of Maude codes are introduced in the next step.

\subsubsection{Formal Verification for Conflict Detection}
With the codes generated in the above step, we use Maude as the formal verification tool to execute the codes for conflict detection. To implement conflict detection in Maude, we follow three steps: device definition, rule definition, and conflict detection. Based on the proposed models for devices, rules, and conflicts, we implement the corresponding steps. The details are as follows:

\begin{enumerate}

\item \textbf{Device Definition.} In Maude, an IoT device is depicted by a tuple $\mathsf{device}(\mathsf{id}, \mathsf{type}, s, \mathsf{location})$. This representation aligns with our device model, with two key differences. First, a Maude device can only have one state at a time. Second, actions are implicitly represented by state transitions, such as $\mathsf{device} \Rightarrow \mathsf{device}'$. We also define a $\mathsf{HomeState}$ to track the collective state of all operated devices.

\item \textbf{Rule Definition.} We define automation rules, denoted as $R_i$, within the Maude framework. A rule that is not contingent upon environmental factors is formalized as an unconditional rewrite rule: 
$$\mathbf{rl}\ [\mathsf{R}]: \mathsf{device1}(), \cdots \Rightarrow \mathsf{device1'()}, \cdots.$$ The $\mathbf{rl}$ operator signifies an unconditional rewrite rule. Conversely, a rule that is influenced by environmental factors is expressed as a conditional rewrite rule:
$$\mathbf{crl}\ [\mathsf{R}]: \mathsf{device1}(), \cdots \Rightarrow \mathsf{device1'()}, \cdots \ \mathbf{If}\ \mathsf{E}.$$
Here, the $\mathbf{crl}$ operator denotes a conditional rewrite rule, and the rule's execution is contingent upon the satisfaction of condition $\mathsf{E}$.

\item \textbf{Conflict Detection.} To detect state conflicts (in $C_S$) and environment conflicts (in $C_E$), we've defined Maude functions, $\mathbf{isIn()}$. These functions check if the current $\mathsf{HomeState}$ contains any conflicting pairs of states during the simulation of automation rules. 
To detect state cascading conflicts and state-environment cascading conflicts, we've defined Maude functions, $\mathbf{isExist()}$. For state cascading conflicts, these functions determine whether the current $\mathsf{HomeState}$ contains $S_i$ of an automation rule $R_i$ during the simulation of the other automation rules. For state-environment cascading conflicts, these functions determine whether the current $\mathsf{HomeState}$ contains $S_i$ and the sensors' measured states in $\mathsf{HomeState}$ satisfy $E_i$.
\end{enumerate}

We utilize \texttt{search} command in Maude to discover the conflicts between the automation rules. Executing the \texttt{search} command returns either a vulnerable state reachable from the initial state, indicating the conflict information donated as $\langle \mathbf{conflict\_information} \rangle$, or no solution, indicating no such conflict. If the result of conflict detection is ``no conflict'', the automation rules in $\langle \mathbf{rule\_list} \rangle$ are conflict-free and would be deployed into the IoT platform to provide convenience for users. Otherwise, the automation rules are optimized in the next step.

\begin{table}[!htbp]  
\centering  
\caption{Prompt $\mathcal{D}$: Automation Rule Optimization}\label{tab:rule_op}  
\begin{tabular}{p{0.12\textwidth}p{0.3\textwidth}} 
\toprule  
\textbf{Elements} & \textbf{Contents} \\  
\midrule  
\textbf{Instruction} &
Modify the automation rules in the input data according to Context, ensuring that the re-output automation rules do not have related conflicts.\\ 
\midrule  
\textbf{Context} & $\langle \mathbf{conflict\_information} \rangle$.
\\  
\midrule  
\textbf{Input data} & $\langle \mathbf{rule\_list} \rangle$. \\  
\midrule  
\textbf{Output indicator} & $\langle \mathbf{new\_rules} \rangle$ in the JSON format:\{``rule1'': \{``trigger'': ``xxx'', ``action'': ``xxx''\} ... \}. \\  
\bottomrule  
\end{tabular}  
\end{table} 

\subsubsection{Rule Optimization}

Using the conflict information produced by Maude, we write a prompt to optimize the automation rules to avoid the conflicts, as shown in Table \ref{tab:rule_op}. The instruction of the prompt is modifying the automation rules in the $\langle \mathbf{rule\_list} \rangle$ according to Context, ensuring that the re-output automation rules do not have related conflicts. We add $\langle \mathbf{conflict\_information} \rangle$ to the context and require that the output $\langle \mathbf{new\_rules} \rangle$  in a JSON format.

\section{Performance Evaluation} ~\label{sec:evaluation}

We have developed a prototype of AutoIoT with 1000+ lines of Python code on top of open source tools such as Maude \cite{Maude} for formal verification and SOTA LLMs for generation tasks. In this section, we present the evaluation results. We first describe our experimental setup in section \ref{Implementation}, and then AutoIoT's performance in section \ref{Performance}. Finally, we demonstrate an AutoIoT assisted smart home case study in section \ref{case study}.

\subsection{Experiment Setup}~\label{Implementation}
\subsubsection{Dataset} We deployed 10 types of IoT devices in our lab on Mi Home platforms. We captured 20 photos for each type of device, totaling 200 photos, to serve as the dataset for device information extraction. The labels of the dataset include bedside lamps, desk lamps, door and window sensors, gateway, human body sensors, motion sensors, pendant lights, smart cameras, smart speakers, and temperature and humidity sensors. We use the hold-out method to randomly divide the data set into the training set and the test set according to the ratio of 7:3.

\subsubsection{Implementation} We selected four popular LLMs to test their performance for device information extraction, rule generation, and formal code generation, including Qwen2.5 \cite{tongyi}, iFLYTEK Spark 4.0 \cite{spark}, Dou Bao \cite{doubao}, and ChatGPT-4o \cite{chatgpt}. We performed our CNN model and executed the formal verification codes on MacOS Sequoia 15.0.1, Apple M1 Pro CPU, 32GB RAM, and 1TB SSD.

\subsection{Evaluation Results}~\label{Performance}
\subsubsection{Time cost of AutoIoT}
This experiment is to test the time cost for different steps of AutoIoT, i.e., device information extraction, rule generation, code generation, and conflict detection, with respect to a variable number of IoT devices. The time cost refers to the time delay between sending a request and getting the result. 

\begin{figure}[!htbp]
    \centering
    \centerline{\includegraphics[width=0.5\textwidth]{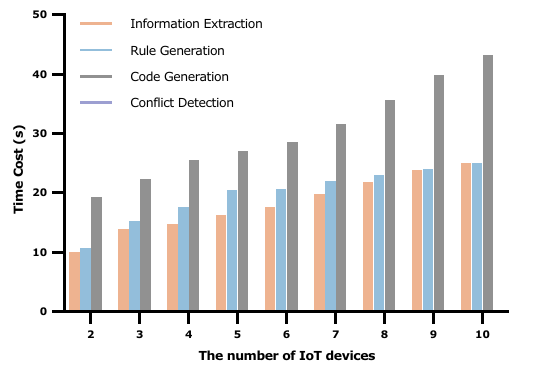}}
	\caption{The time cost for device information extraction, rule generation, code generation, and conflict detection with different numbers of IoT devices.} 
	\label{fig:test1}
\end{figure}

We first trained a ResNet-18 model to extract the device information using the dataset that we made. After that, we took 10 photos containing varying numbers of IoT devices and their manuals as the inputs for this experiment. The LLM we used in this experiment is Qwen2.5 because its performance is the best of the four LLMs in our experiment. As shown in Fig. \ref{fig:test1}, one can see that the time cost for each step except conflict detection is at the second level, and it increases with the increasing number of IoT devices, while the time cost for conflict detection is at the millisecond level. The reason for the aforementioned phenomenon is that conflict detection is executed locally, consuming less time, whereas the other three stages involve generation tasks that require the participation of LLM in the execution process, leading to a significant increase in time consumption. The time cost of code generation is the longest, roughly from 19.6s to 42.4s, which is reasonable because code generation involves higher structural complexity than natural language and requires a deeper understanding of context.

\subsubsection{Time cost with different LLMs}
This experiment is to test the time cost associated with using different LLMs (i.e., Qwen2.5 \cite{tongyi}, iFLYTEK Spark 4.0 \cite{spark}, Dou Bao \cite{doubao}, and ChatGPT-4o \cite{chatgpt}) for various generation tasks of AutoIoT, including device information extraction, rule generation, and code generation. We took one photo containing five IoT devices as the input of AutoIoT. The five devices consist of a desk lamp, a door and window sensor, a smart camera, a motion sensor, and a bedside lamp. 
Based on the raw data obtained from the experiment, we found that the time cost of different large models is difficult to compare due to varying network latencies when invoking these LLMs. Therefore, we only compared the time cost ratios of different LLMs for processing each task, as shown in Table~\ref{tab:test2}. The time cost for code generation occupies the largest proportion because generating is not merely a matter of simple text concatenation. It involves complex logic construction, syntactic correctness, variable management, and the understanding and application of advanced concepts such as loops and conditional statements. To ensure that the generated code is both effective and efficient, LLMs need to perform extensive computations to evaluate various possibilities.
One can see that the time cost ratios for the same task of different LLMs are different. The time cost ratio for code generation of Qwen 2.5 is the smallest of the four LLMs, while the ratio of Dou Bao is the biggest. The reason for this may be the differences in the model architecture of LLMs.

\begin{table}[!ht]
\centering
\caption{The time cost ratio of information extraction, rule generation, and code generation with different LLMs}\label{tab:test2}
\begin{tabular}{cccccc}
\toprule
\textbf{LLMs}   & \textbf{\begin{tabular}[c]{@{}c@{}} Information\\ Extraction \end{tabular} } & \textbf{Rule Generation} & \textbf{Code Generation} \\
\midrule
 Qwen 2.5 & 25.06       &  31.14     &  43.80   \\
\midrule
\begin{tabular}[c]{@{}c@{}}iFLYTEK \\ Spark 4.0\end{tabular}  &  17.06    & 21.67       &  61.27   \\
\midrule
Dou Bao &   13.76      &  18.36     & 67.88  \\
\midrule
ChatGPT-4o & 14.90 & 32.60 & 52.50 \\
\bottomrule
\end{tabular}
\end{table}

\subsection{Case Study: AutoIoT Assisted Home Automation} ~\label{case study}
We report a case study that employs AutoIoT to assist smart home automation. This case study shows how to integrate AutoIoT into popular IoT platforms.

In smart home systems, automation rules are typically authored by users in trigger-action format based on their understanding and needs. This raises the barrier to user experience with automation, as it requires users to have a certain level of knowledge regarding smart home automation. If pre-defined automation rules are available, users can simply add these rules to their smart home systems to experience automation. This may appear to be a good solution and in fact it has been adopted by popular smart home platforms such as Mi Home \cite{mihome}. However, adding automation rules through templates allows users to experience only a limited set of fixed automation rules, which may not meet their specific needs. Additionally, conflicts may exist among the automation rules, and smart home platforms cannot detect such conflicts before rule execution, potentially leading to safety issues. In this section, we show that with AutoIoT, one can produce the automation rules without knowledge of smart home systems and have high confidence that there are no conflicts among the generated automation rules.

\begin{figure}[!htbp]
    \centering
    \centerline{\includegraphics[width=0.49\textwidth]{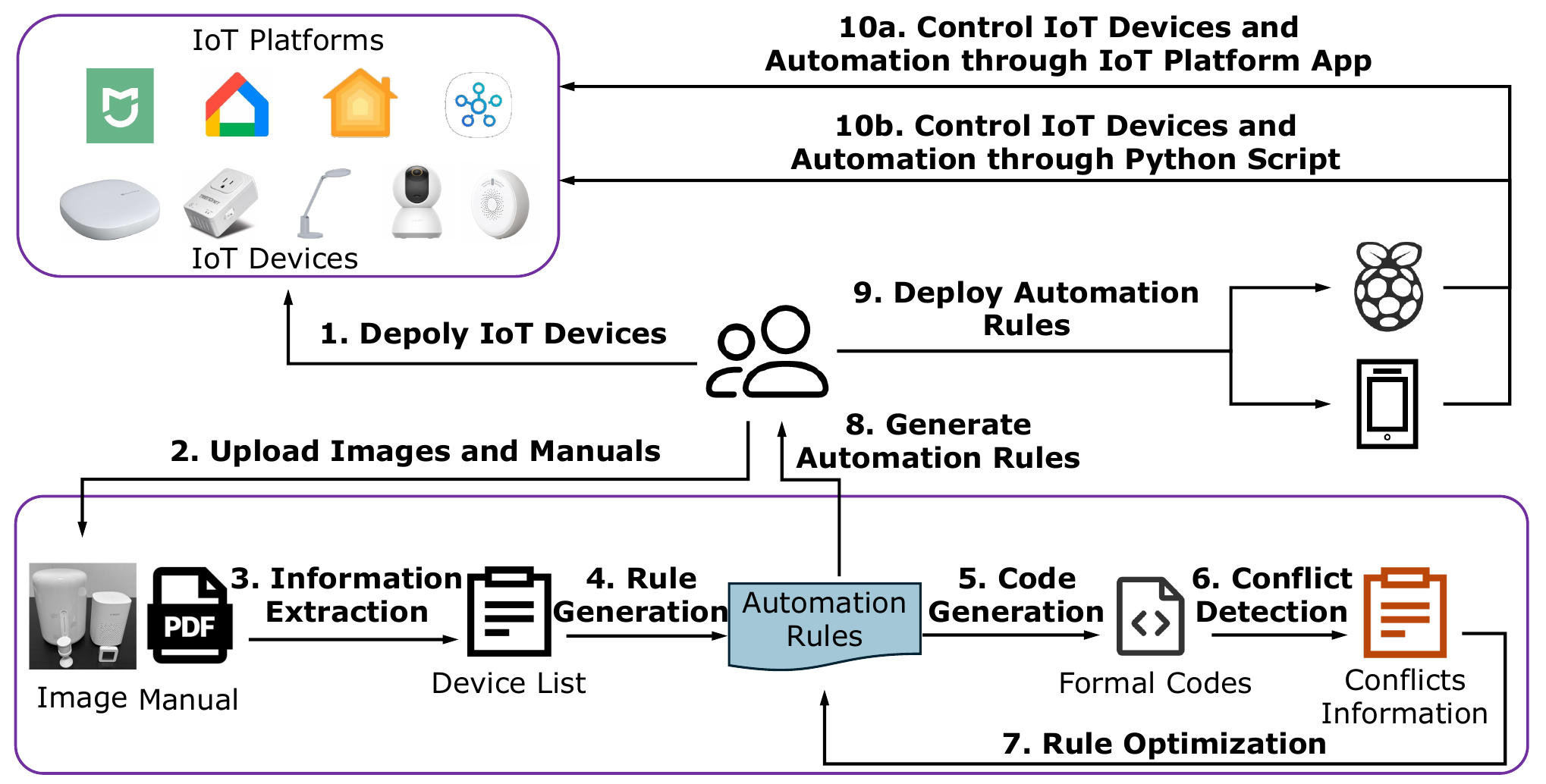}}
	\caption{AutoIoT Assisted Smart Home.} 
	\label{fig: case study}
\end{figure}

Fig.\ref{fig: case study} demonstrates our AutoIoT assisted smart home case study. When a user buys some IoT devices, the user first adds the devices to the corresponding IoT platform through its app. To get some safe automation rules, the user takes photos of the devices and sends the photos, the manuals of each device, and the user's ideas to AutoIoT. Then, AutoIoT would process the user's input according to the workflow depicted in Fig. \ref{fig: architecture} and output a series of safe automation rules that meet the user's requirements. The automation rules consist of two types to cater to the diverse needs of different users. The first type is the automation rules described in natural language in a Trigger-Action format, which users can manually add to the corresponding app to achieve automation. The second type is the automation rules in Python script format, using the python-miio packege\footnote{https://github.com/rytilahti/python-miio}, which users can save and run in their local environment to implement automation. In the following, we present this case study in three processes.

\begin{figure}[!htbp]
    \centering
    \centerline{\includegraphics[width=0.49\textwidth]{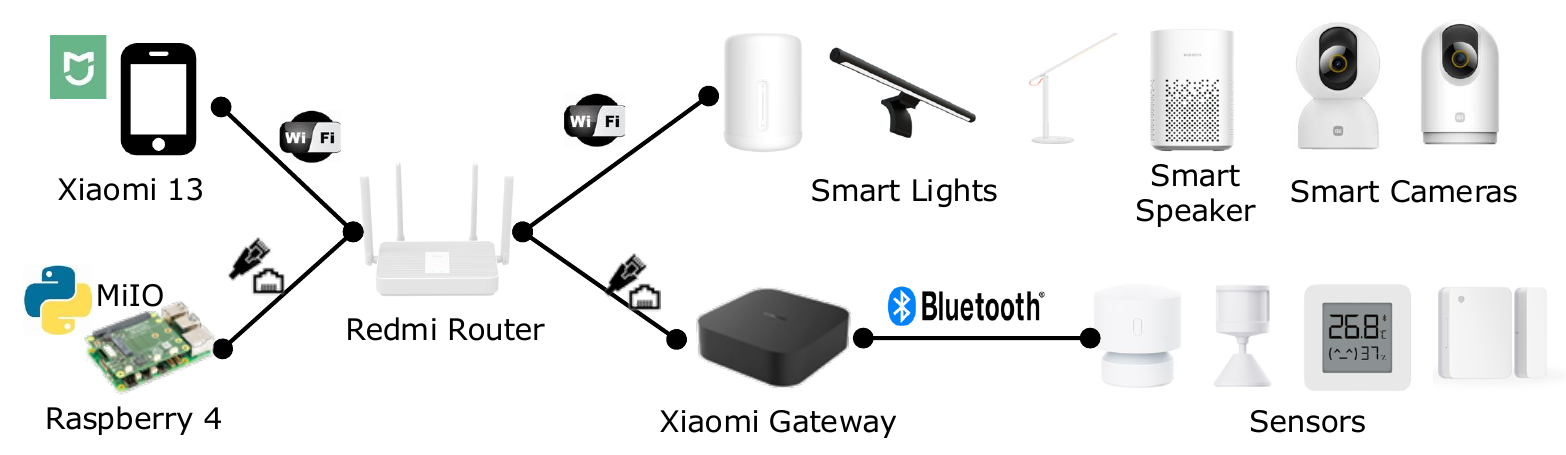}}
	\caption{The device deployment topology of the case study.} 
	\label{fig: deploy}
\end{figure}

\textbf{IoT Devices Deployment.} To establish the smart home for our case study, we deployed 10 Mi Home IoT devices, as is shown in Fig. \ref{fig: deploy}, including one gateway for device connectivity and control, four sensors for detecting environmental changes, two smart cameras, one smart speaker, and three types of smart lights. We placed the smart speaker, bedside lamp, and human body sensor in the bedroom, while the other devices were deployed in the living room. In this setup, all the sensors connect to the gateway via Bluetooth to join our smart home, while other devices join the smart home through Wi-Fi.

\begin{figure}[!ht]
    \centering
    \centerline{\includegraphics[width=0.49\textwidth]{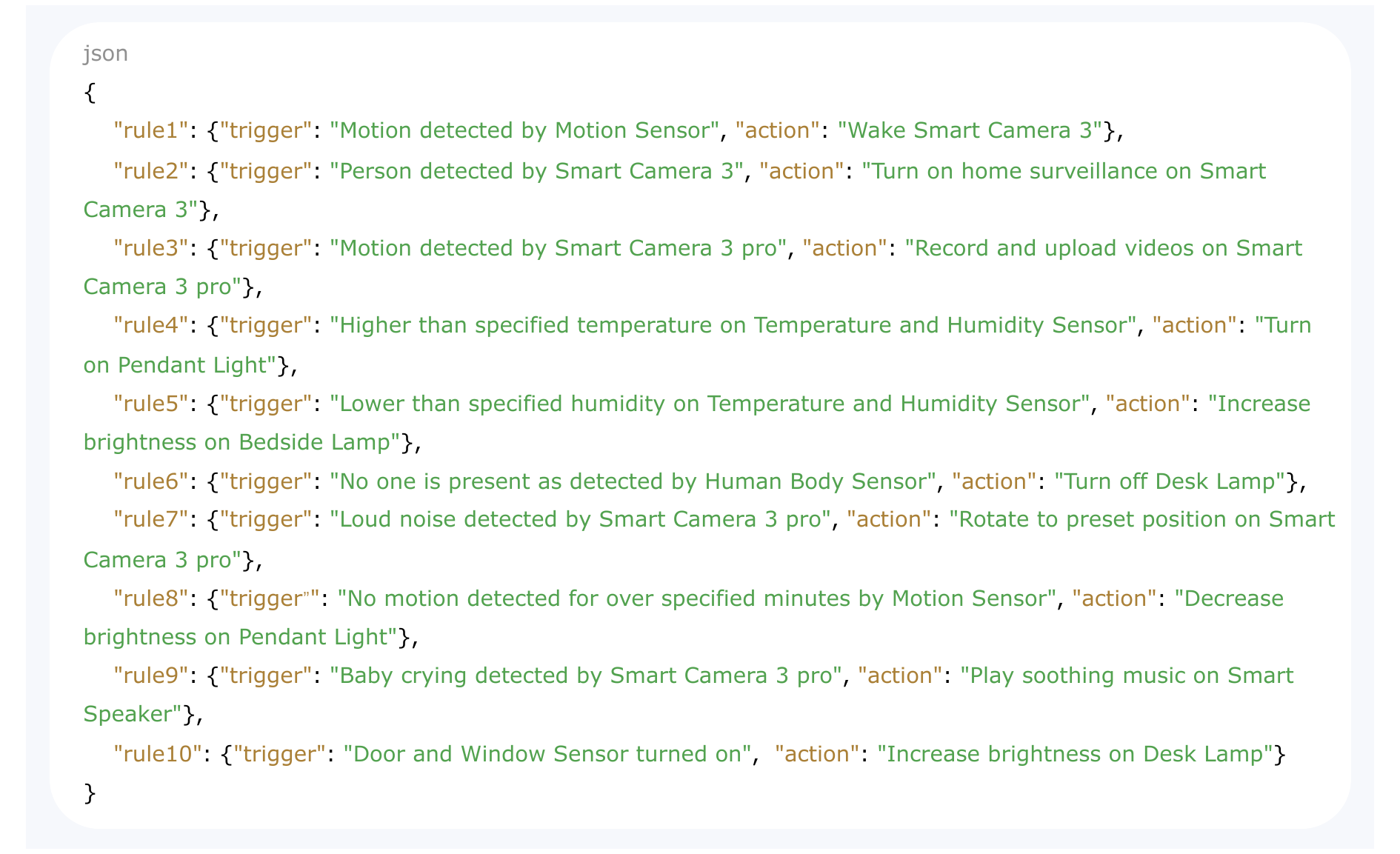}}
	\caption{10 automation rules generated by LLM.} 
	\label{fig:cs2}
\end{figure}

\textbf{Conflict-free Automation Rules Generation.} We first captured a photo and uploaded the photo and 10 manuals as the inputs of AutoIoT. To recognize the IoT devices in photos, we made a dataset of 10 IoT devices, consisting of a smart camera 3, a smart camera 3 pro, a desk lamp, a pendant light, a bedside lamp, a smart speaker, a temperature and humidity sensor, a motion sensor and a human body sensor. Then we used the dataset to train a ResNet-18 model to recognize devices. After that, AutoIoT generated a JSON list of the IoT devices. Then, AutoIoT automatically added the list to the predefined prompt in Table \ref{tab:rule_gen} and interacted with Qwen 2.5 to generate 10 automation rules, as shown in Fig. \ref{fig:cs2}. Based on the rules, LLM-LoT produced the formal verification logic in the format of a code generation template and converted the logic into Maude codes through the code generation adapter. Using the Maude as a formal verification tool, AutoIoT executed the Maude codes and separately checked for the existence of direct state conflict, state impact conflict, direct environment impact conflict, or indirect environment impact conflict among the automation rules by using \texttt{search} command. The results of all conflict detection are no solution, indicating no conflicts among the rules. Thus, the generated rules are conflict-free automation rules.

\textbf{Automation Rules Deployment.} After receiving the automation rules, we tried two deployment schemes to add them to the Mi Home platform. As shown in Fig. \ref{fig: case study}, for the rules in natural language format, we manually added the rules to the automation through the Mi Home app one by one. For the rules in Python script format, we wrote the IP addresses and tokens of the IoT devices into the scripts and ran the scripts on a Raspberry Pi 4 with Ubuntu 24.04 LTS, thereby achieving automated control of the IoT devices.   

\begin{figure}[!h]
    \subfloat[]{\includegraphics[width=0.15\textwidth]{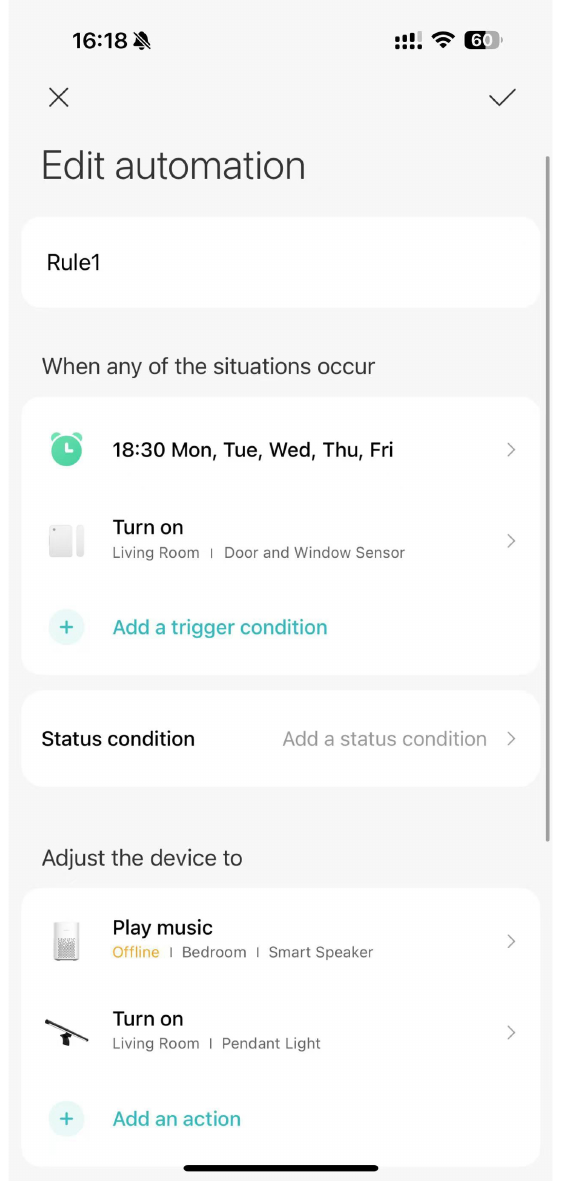}}
  \subfloat[]{\includegraphics[width=0.336\textwidth]{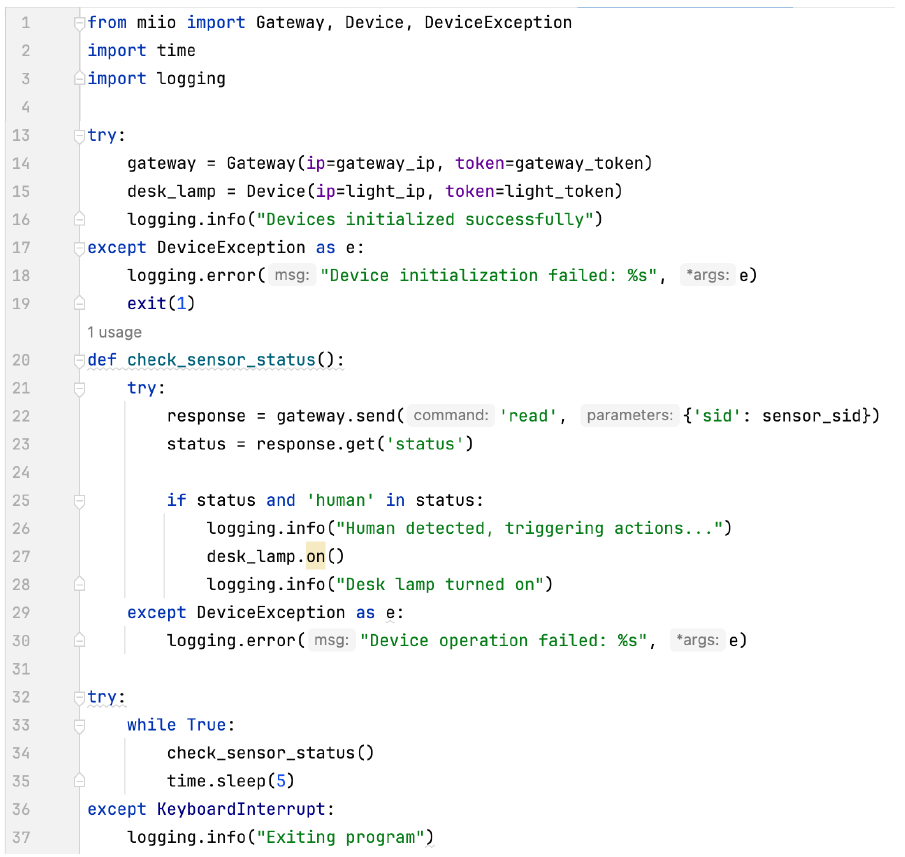}}
  \caption{Examples of deploying the automation rules. (a) Using Mi Home app. (b) Using Python script.} 
  \label{fig:photo}
\end{figure}

\section{Conclusion and Future Directions} ~\label{sec:conclusion}
In this paper, we propose an automated IoT platform called AutoIoT, which is based on LLMs and formal verification techniques, designed to help users generate conflict-free automation rules and assist developers in generating codes for conflict detection, thereby enhancing their experience. We design four prompts for AutoIoT to guide LLMs in performing device information extraction, rule generation, code generation, and rule optimization. We present a code adapter to separate logical reasoning from the syntactic details of code generation, enabling LLMs to generate code for programming languages beyond their training data. Finally, we implement and evaluate AutoIoT, and present a case study to demonstrate how AutoIoT is integrated into popular IoT platforms.
In the future, we plan to expand the smart home device dataset we have created by incorporating a wider variety of smart home devices, allowing us to conduct large-scale testing of AutoIoT. Additionally, we intend to explore the automated generation of other types of policies or codes, such as access control policies, further enhancing the capabilities and utility of our platform.

\section{Acknowledgement}
This study was partially supported by the National Natural Science Foundation of China (No. 62302266, 62232010, U23A20302), the Shandong Science Fund for Excellent Young Scholars (No.2023HWYQ-008), the project ZR2022ZD02 supported by Shandong Provincial Natural Science Foundation.

\bibliographystyle{IEEEtran}
\bibliography{references}

\end{document}